# Tunable Dirac interface states in topological superlattices

G. Krizman[1,2], B.A. Assaf[1], T. Phuphachong[2], G. Bauer,[3] G. Springholz,[3]
G. Bastard,[2] R. Ferreira,[2] L.A. de Vaulchier,[2] and Y. Guldner[2]

[1] Département de Physique, Ecole Normale Supérieure, Paris Sciences et Lettres Research University, CNRS, 24 rue Lhomond, 75005 Paris, France

[2] Laboratoire Pierre Aigrain, Département de Physique, Ecole Normale Supérieure, Paris Sciences et Lettres Research University, Sorbonne Université, CNRS, 24 rue Lhomond, 75005 Paris, France

[3] Institut für Halbleiter- und Festkörperphysik, Johannes Kepler Universität, Altenberger Straβe, 69, 4040 Linz, Austria

## Abstract

Relativistic Dirac fermions are ubiquitous in condensed-matter physics. Their mass is proportional to the material energy gap and the ability to control and tune the mass has become an essential tool to engineer quantum phenomena that mimic high energy particles and provide novel device functionalities. In topological insulator thin films, new states of matter can be generated by hybridizing the massless Dirac states that occur at material surfaces. In this paper, we experimentally and theoretically introduce a platform where this hybridization can be continuously tuned: the $Pb_{1-x}Sn_xSe$ topological superlattice. In this system, topological Dirac states occur at the interfaces between a topological crystalline insulator $Pb_{1-x}Sn_xSe$ and a trivial insulator, realized in the form of topological quantum wells (TQW) epitaxially stacked on top of each other. Using magnetooptical transmission spectroscopy on high quality molecular beam epitaxy grown $Pb_{1-x}Sn_xSe$ superlattices, we show that the penetration depth of the TQW interface states and therefore their Dirac mass is continuously tunable with temperature. This presents a pathway to engineer the Dirac mass of topological systems and paves the way towards the realization of emergent quantum states of matter using $Pb_{1-x}Sn_xSe$ topological superlattices.

# I. Introduction

Engineered quantum effects in heterostructures [1] of condensed-matter are behind several revolutionizing ideas and advances in fundamental and applied physics. Examples are the quantum spin Hall [2] and fractal quantum Hall effects [3] in semiconductor and graphene heterostructures, as well as band engineered quantum cascade lasers [4] and entangled photon sources. [5] With the emergence of topological phases of matter and the discovery of topological insulators, [6] [7] a new horizon has been opened to develop heterostructures combining materials of alternating band topology. In topological insulators, Dirac cones emerge at the surfaces of an insulator hosting a bulk band inversion. In a superlattice of alternating topological and trivial materials, these states appear at the topological material's interfaces, and their mutual coupling and hybridization can lead to novel quantum phases such as Weyl fermions [8] and Su-Schriefer-Heeger (SSH) excitations. [9] A natural question is whether and how one can engineer the hybridization between the topological interface states and tune the emergent nontrivial topological phases using external knobs.

A highly attractive material system to achieve this purpose are the topological crystalline insulators (TCI) such as $Pb_{1-x}Sn_xSe$ and $Pb_{1-x}Sn_xTe$. [10] - [16] Topological phase transitions can be induced in TCIs by tailoring the crystal structure and breaking crystalline symmetries. [11] [14] [17] - [20] $Pb_{1-x}Sn_xSe$ and $Pb_{1-x}Sn_xTe$ host four Dirac cones per surface Brillouin zone (Fig. 1(a)). [21] [22] [23] For example, their characteristics – the effective electron mass and energy gap – are highly sensitive to strain [24] [25] and lattice distortions [19] and can be controlled by electric [18] and magnetic fields. [26] This has led to novel theoretical proposals such as the electrical-field-tuned quantum spin Hall state, [18] [27] and the helical interfacial flat band state that appears upon straining those materials. [25] The latter is particularly interesting as it is predicted to occur in heterostructures of TCIs separated by normal insulating barriers, [25] and yields an emergent superconducting phase with an enhanced critical temperature. [28] [29] Heterostructures of TCIs are therefore an exciting platform to engineer tunable topological phases.

Up to now, however, most studies on TCIs dealt with bulk single crystals and thick epilayers, [13] [22] [30] that required surface sensitive and local probes to thoroughly characterize the topological states under ultra-high vacuum conditions far from practical device applications. In particular, these techniques are not sensitive to states occurring at buried interfaces in superlattices. This fact impeded the prospects of realizing novel states of matter in these systems.

In this work, we unravel the topological interface states (TIS) formed in epitaxial TCI superlattices consisting of multiple $Pb_{1-x}Sn_xSe$ topological quantum wells (TQW) separated by normal insulator (NI) $Pb_{1-y}Eu_ySe$ barriers (Fig. 1(b,c)). Using magneto-infrared spectroscopy to probe the Landau level structure of the quantum confined states formed at the buried interfaces, we reveal the behavior of the TIS as a function of temperature. Detailed theoretical analysis of the experimental data reveals that the penetration depth, the hybridization strength and the Dirac mass of the TIS can be effectively tuned not only by varying the quantum well thickness but also by continuously changing temperature (Fig. 1(d)). Our results thus establishes $Pb_{1-x}Sn_xSe$ topological quantum wells (TQWs) as a viable platform that hosts tunable topological

Dirac states at buried interfaces. Our findings are of key importance for the realization of new emergent states of matter in TI and TCI heterostructures. [25] [9]

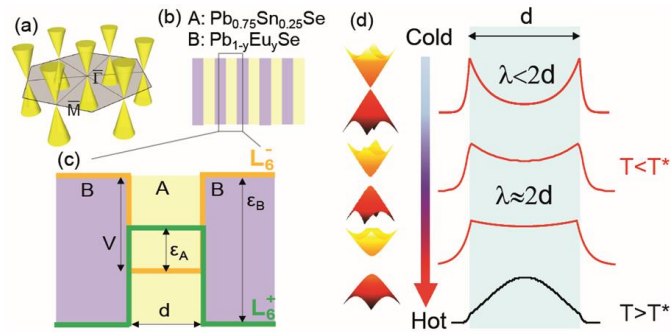

**FIG 1.** (a) Identical Dirac cones on the (111)-surfaces of bulk $Pb_{0.75}Sn_{0.25}Se$. (b) Topological quantum-well $Pb_{0.75}Sn_{0.25}Se/Pb_{1-y}Eu_{y}Se$ superlattices of alternating TCI/NI layers studied in this work. The band alignment and band profiles are shown in (c). $L_6^{\pm}$ denote the conduction and valence band extrema at the L-points in the rocksalt Brillouin zone of $Pb_{0.75}Sn_{0.25}Se$ and $Pb_{1-y}Eu_{y}Se$. $\varepsilon_A$ (<0 at 4.2K) is the $L_6^{\pm}$ energy separation in the band-inverted $Pb_{0.75}Sn_{0.25}Se$ quantum wells (A) and $\varepsilon_B$ (>0) that of the normal insulator $Pb_{1-y}Eu_{y}Se$ barriers (B). V is the conduction band offset and *d* the well thickness. (d) Sketch of the evolution of the wave function probability density and Dirac cones of the topological interface state as a function of temperature across the topological phase transition T*, illustrating the tunability of the penetration depth and hybridization gap of the TIS.

## II. Experimental method

Magneto-optical spectroscopy experiments [31] [32] are performed on coherent TCI superlattices of $Pb_{0.75}Sn_{0.25}Se/Pb_{1-y}Eu_{y}Se$ grown by molecular beam epitaxy on (111) $BaF_2$ substrates (see Appendix A). The Sn content x = 0.25 of the TQW was chosen to be above the TCI phase transition that occurs at x = 0.16 at T=4.2K. [32] Infrared spectroscopy is performed at different temperatures and magnetic fields on two TQW samples. The first – TQW 36 – has a quantum and barrier thickness of 36nm and 26 periods. The second – TQW 24 – has 24nm quantum wells, 120nm barriers and 16 periods.

Spectra were recorded in an optical cryostat equipped with a superconducting magnet providing magnetic fields up to 15T and temperatures down to 4.2K. The probe is coupled to a FTIR spectrometer that operates in the far-IR and mid-IR to cover a range extending from roughly 1meV to 1eV. The signal was either recovered at the exit of the probe and measured using an external HgCdTe detector cooled to 77K or detected inside the probe using a composite Si bolometer mounted below the sample and cooled to 4.2K. The external detection setup was used to perform temperature dependent measurements between 4.2K and 200K. The detection cutoff in this case is close to 80meV. All the measurements are performed in the Faraday geometry with B || [111] direction, the growth axis. It is emphasized that due to the large penetration depth of the infrared photons, the entire heterostructure is probed.

## III. Results
### A. *Magnetooptical specotroscopy of topological interface states*

Figure 2(a) shows relative transmission spectra acquired at different magnetic fields for sample TQW-36 at 4.2K. The spectra exhibit pronounced minima due to interband Landau level (LL) transitions that are observed at magnetic fields down to 2T. Their strength significantly gains in amplitude and strongly blue-shifts with the applied magnetic field. It is therefore obvious that they originate from Landau quantization. The large number of transitions evidences the high mobility and low carrier density of the TQW structures resulting from the pseudomorphic heteroepitaxial growth and effective control of the carrier concentration, [32] with a Fermi energy less than 40meV above the mid-gap. Corresponding spectroscopic data for TQW-24 are shown in the supplemental material. [33]

From the magneto-optical spectra, we construct fan charts of the Landau level transitions in Fig. 2(b) to extract the band structure parameters. As a first step, we identify the transitions of the TIS (E1 and H1) marked by the red arrows in Fig. 2(a) and represented in (b) by red points. A second set of weaker transitions also appears (black arrows in Fig. 2(a)) and is assigned to transitions between the second subbands E2-H2 and hybrid ones between TIS and E3 (see Fig. 2(c)). These are represented by the open black and purple circles in the fan chart. Using the k·p envelope function approach developed in Refs. [1], [34], [35] and detailed in appendix B, we compute the quantum confined states and magneto-optical transitions, taking into account the opposite topological character of the TQWs and NI barriers (Fig. 1(c)). The fits represented by the solid lines in Fig. 2(b) evidence an excellent agreement with the experimental data. This yields the material band structure parameters listed in Table I.

**Table I**. Structure and band parameters of the TCI/NI TQW-36 and TQW-24 investigated in this work. $d$ and $d_{barrier}$ are the $Pb_{1-x}Sn_xSe$ and $Pb_{1-y}Eu_ySe$ thicknesses and x and y their composition as obtained by x-ray diffraction (see appendix 1). The bulk band gaps $\varepsilon_A$ and $\varepsilon_B$ of the TQW and barriers as well as the Dirac velocity $v_c$ derived from Landau level spectroscopy and k·p analysis are also listed (see supplement S2 and S3). [33] [36]' [37]' [38] [39]

| Sample | $Pb_{1-x}Sn_xSe$ TQW | $d$ (nm) | $Pb_{1-y}Eu_ySe$ barrier | $d_{barrier}$ (nm) | # of periods | $\varepsilon_A$, (4.2K) (meV) | $\varepsilon_B$ (4.2K) (meV) | $v_c$ (m/s) |
|---|---|---|---|---|---|---|---|---|
| TQW-36 | x=0.25 | 36 | y=0.056 | 36 | 26 | -40 | 314 | $4.5 \times 10^5$ |
| TQW-24 | x=0.25 | 24 | y=0.04 | 120 | 16 | -40 | 270 | $4.6 \times 10^5$ |

The $E(k)$ dispersion of the TQW states and their Landau levels versus magnetic field derived from the k·p fits are presented in Fig. 2(c) and (d), respectively. The calculated LL transition curves (solid lines in Fig. 2(b)) are obtained by computing the magneto-optical interband transitions indicated by the arrows in Fig. 2(d), taking into account optical selection rules that mix parity ( $\Delta i$ = 0,2 where $i$ denotes the band index, and $\Delta N=\pm 1$, where $N$ is the Landau level index). Notice that the strongest transitions highlighted in red in Fig. 2(a,b,d) are Dirac dispersing with a characteristic $\sqrt{B}$ dependence of the LL levels. They are due to a Dirac state with a very small energy gap Δ=5meV shown in Fig. 2(c). Corresponding theoretical calculations of its probability density (inset of Fig. 2(c)) demonstrate that this Dirac state is pinned to the interfaces. Thus, the E1 and H1 states correspond to the TIS of the $Pb_{1-x}Sn_xSe$ quantum wells. Our analysis reveals the existence of a small Dirac energy gap of Δ=5meV of the TIS interface states caused by the hybridization of the top and bottom TIS, the wavefunctions of which exhibit a finite overlap shown in Figure 2(c). This effect is a hallmark of quantum confined topological states in TCIs, observed here for buried TCI/NI quantum well heterostructures. We have elucidated this effect using Landau level magnetospectroscopy, whereas previously the hybridization gap could only be observed at the free surface of extremely thin TI or TCI films. [16] [40] [41]

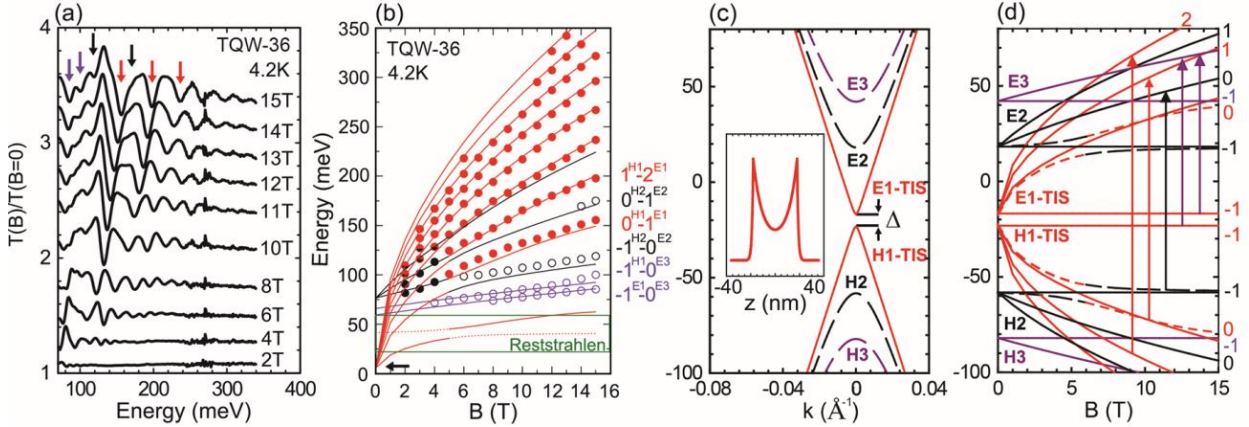

**FIG 2.** (a) Relative transmission spectra ($T(B)/T(0)$) versus photon energy at 4.2K for magnetic fields between 2T and 15T. The curves are shifted for clarity. (b) Landau fan chart derived from the experiments and k·p envelope function calculations (solid lines). Red data points and lines are the TIS (E1-H1) interband transitions, black points and line are the H2-E2 transitions and the purple points and lines denote hybrid transitions between the TIS and E3. The green rectangle indicates the reststrahlen region of the BaF$_2$ substrate that blocks the infrared transmission. (c) Calculated k·p subbands and dispersion of TQW-36. The inset shows the probability density ($\chi$) of the E1 state, i.e., the TIS at 4.2K. (d) Computed Landau levels versus magnetic field. The lowest LL is denoted by N=-1 for each subband. The dashed lines are computed by taking into account an anti-crossing (see appendix 4) of 5meV between $-1^{E2/H2}$ and $0^{E1/H1}$ levels. The red arrows in (a) and (d) indicate the allowed magneto-optical transitions between the TIS (E1-H1), the black and purple arrows indicate the transition between the second QW subbands (E2-H2) and between the TIS and E3, respectively.

From the measured energy gap $\Delta$ and Dirac velocity $v_c$, [32] [42] we can compute the Dirac mass of the TIS:

$$m_D = \frac{\Delta}{2v_c^2} = 0.0022 m_0$$

at 4.2K. Indeed the mass acquisition due to hybridization is extremely small since the QW is relatively large (36nm), evidencing nearly-massless Dirac states at the TCI/NI interface. It is important to note at this point, that while for SnTe and Pb$_{1-x}$Sn$_x$Te (111) two types of Dirac cones have been observed at the surface, [43] in the case of Pb$_{1-x}$Sn$_x$Se, the bulk Fermi surface is nearly-isotropic for x≈0.25 as reported previously. [32] [44] Consequently, although Dirac cones occur at two different crystalline symmetric points of the Brillouin zone ($\bar{\Gamma}$ and $\bar{M}$) as shown in Fig. 1(a), their velocity and their shape are identical. The measured Dirac mass thus corresponds to all 4 Dirac valleys of Pb$_{1-x}$Sn$_x$Se.

A remarkable feature of our result is that the amplitude of the magneto-optical transitions from the Dirac interface states is larger than those stemming from the trivial QW subbands. This means that in heterostructures the TIS transitions dominate the signal, contrary to the case of bulk material. We will next reveal a peculiarity of this TIS of $Pb_{1-x}Sn_xSe$ – its temperature tunability.

### B. Tuning the topological interface states with temperature

For topological $Pb_{1-x}Sn_xSe$ one can tune the magnitude of the bulk energy gap $\varepsilon_A$ of the system by changing the temperature. [45] [46] Hence, one can vary the penetration depth $\lambda$ of the TIS into the $Pb_{1-x}Sn_xSe$ well according to (see appendix 2):

$$\lambda(T) = \frac{2\hbar v_c}{\sqrt{\varepsilon_A(T)^2 - \Delta(T)^2}} \quad (1)$$

Eq. (1) is equivalent to the expression of $\lambda$ derived in previous works on TI and TCI thin films. [47] [48] Its temperature dependence is a hallmark of the tunability of TCIs and has not yet been considered or observed. The variation of $\lambda$ related to the changing Dirac mass of charge carriers, $\lambda$ versus $T$ can therefore be traced by measuring the change in $m_D$ versus $T$.

| T(K) | $\varepsilon_A$ (meV) | $v_c$ (m/s) |
|---|---|---|
| 4.2 | -40 | 4.5x10$^5$ |
| 20 | -35 | 4.5x10$^5$ |
| 40 | -30 | 4.5x10$^5$ |
| 60 | -20 | 4.5x10$^5$ |
| 80 | -12.5 | 4.5x10$^5$ |
| 120 | 5 | 4.5x10$^5$ |
| 160 | 20 | 4.5x10$^5$ |
| 200 | 35 | 4.5x10$^5$ |

Table II. Model parameters and TIS gap for TQW-36 at all studied temperatures. See supplement (S2) for $\varepsilon_B(T)$.

The experimental results shown in Fig. 3 for TQW-36 demonstrate how we achieve a tuning of $m_D$ and a controlled suppression of TIS via temperature by tailoring $\lambda$ for a fixed QW thickness. Remarkably, the sample mobility is sufficiently high to resolve the LLs at temperatures as high as 200K as evidenced in Fig. 3(a). Figure 3(b-e) show the resulting Landau level fan charts at different temperatures $T$ = 60K, 80K, 120K and 200K; the results for TQW-24 are shown in the supplement. [33] The whole experimental data set is fit by the k·p model that takes into account the variation of $\varepsilon_A$ versus $T$ obtained from a bulk-like reference sample. [33] The model parameters are shown in table II. The excellent fit to the data is shown by the solid lines in Fig. 3(b-e).

From the theoretical analysis of the experiments, we extract the temperature dependent Dirac gap Δ(T) (arrows in Fig. 3(b-e)) and the corresponding Dirac mass $m_D(T)$. They are plotted in Fig. 3(f) for both TQW samples. A monotonic increase of Δ and $m_D$ is seen with increasing temperature, whereas the bulk gap $\varepsilon_A$ changes from *negative to positive* at 110 K. Evidently, for the thinner TQW-24, Δ is systematically larger, which unambiguously confirms the quantum confinement and hybridization related nature of the gap, and the topological character of the TQW states.

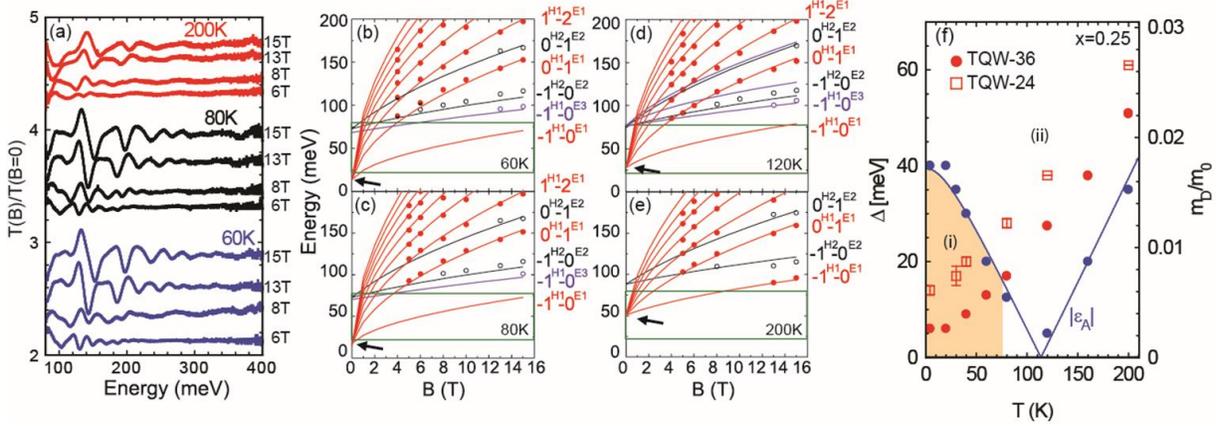

**FIG 3.** (a) Relative transmission spectra (T(B)/T(0)) for TQW-36 recorded at T = 60K – 200 K for four magnetic fields from 6T to 15T. The curves are shifted vertically for clarity. (b-e) Resulting Landau level fan charts for T = 60, 80, 120 and 200K. The experimental data is shown as red dots, black and purple circles for the E1-H1, respectively, E2-H2 and the hybrid H1-E3 and E1-E3 transitions, and the fit by k·p theory is represented by the solid lines. The black arrows mark the E1-H1 Dirac gap Δ. The green rectangle indicates the reststrahlen region of the $BaF_2$ substrate that blocks the infrared transmission. (f) Derived hybridization gap Δ (left axis) and Dirac mass $m_D$ (right axis) of the TIS versus temperature for TQW-36 (red dots) and TQW-24 (red open squares). The variation of the bulk gap |$\varepsilon_A$| versus temperature for $Pb_{1-x}Sn_xSe$ with the given composition (x = 0.25) is shown by blue dots and solid line. The orange shaded region represents the topological regime where Δ < |$\varepsilon_A$| for TQW-36.

## C. Discussion

Two regimes (i) and (ii) emerge from Fig. 3(f) when the Dirac gap Δ is compared to the bulk gap $\varepsilon_A$. *Regime (i)* occurs when Δ<|$\varepsilon_A$| and $\varepsilon_A$ < 0, i.e., when the Dirac gap of the TIS is smaller than the bulk gap that is also inverted. According to Fig. 3(f) this applies when T is below the critical temperature T*=70 K for TQW-36, and for T<T*=50K for TQW-24. In the *second regime (ii)*, Δ>|$\varepsilon_A$|. To shed light on the nature of the Dirac states in these regimes, we evaluate the probability density obtained by the k·p calculations based on the fit to the data.

Figure 4(a) and 4(b) display the results for TQW-36 at T=4.2K and 60K (<T*) for regime (i). The band alignment diagram and E(k) dispersions reveal that the E1 and H1 TQW states (red lines) lie within the $Pb_{1-x}Sn_xSe$ bulk gap (Δ<|$\varepsilon_A$|). As a result, the probability density χ of the Dirac states is peaked at the interfaces (red solid lines in Fig. 4(a,b)). Its decays towards the center of the quantum well is governed by λ given by Eq. (1). Approaching the critical temperature T* where Δ→|$\varepsilon_A$|, the TIS penetrates deeper into the QW (see Fig. 4(b) at 60K). As a result, the coupling strength and hybridization gap Δ of the TIS increases (see

Fig. 3(f)), even though $Pb_{1-x}Sn_xSe$ is still in the non-trivial TCI state with $\varepsilon_A<0$. Therefore, the observed increase of $\Delta$ is accompanied by an increasing penetration $\lambda$ of the TIS as shown in Fig. 4(e). Through this relation, a tuning of both $\lambda$ and $m_D$ is achieved for the TIS.

A key feature of this regime is that as long as $\Delta <|\varepsilon_A|$, the ground state remains peaked at the interface and is of topological origin. This suggests that topological characteristics such as spin-momentum locking are still partially preserved even when $\Delta\neq 0$. Previous ARPES measurements on $Bi_2Se_3$ films in fact observed a progressive suppression of the spin-polarization of the surface states upon gapping, [40] suggesting that upon increasing top-bottom surface hybridization the loss of spin-momentum locking is only partial as long as the states remain pinned to the surface or interface.

Further increasing the temperature leads to a transition into regime (ii) where $\Delta >|\varepsilon_A|$ (with either $\varepsilon_A>0$ or $\varepsilon_A<0$). As shown by Fig. 4(c,d) for TQW-36 at T=80K and T=200K (>$T^*$), the E1 and H1 energy levels move, respectively, above and below the bulk TCI band edges. As revealed by the computed probability density $\chi$ displayed in Fig. 4(c,d), the wave function of E1/H1 is no longer peaked at the interfaces. Here, $\Delta$ is simply the confinement induced gap between the first E1/H1 QW subbands. Note that the interfacial nature of E1 and H1 remains unchanged when $\varepsilon_A$ changes sign (at 110K), as it is entirely determined by the value of $\Delta$ relative to $|\varepsilon_A|$.

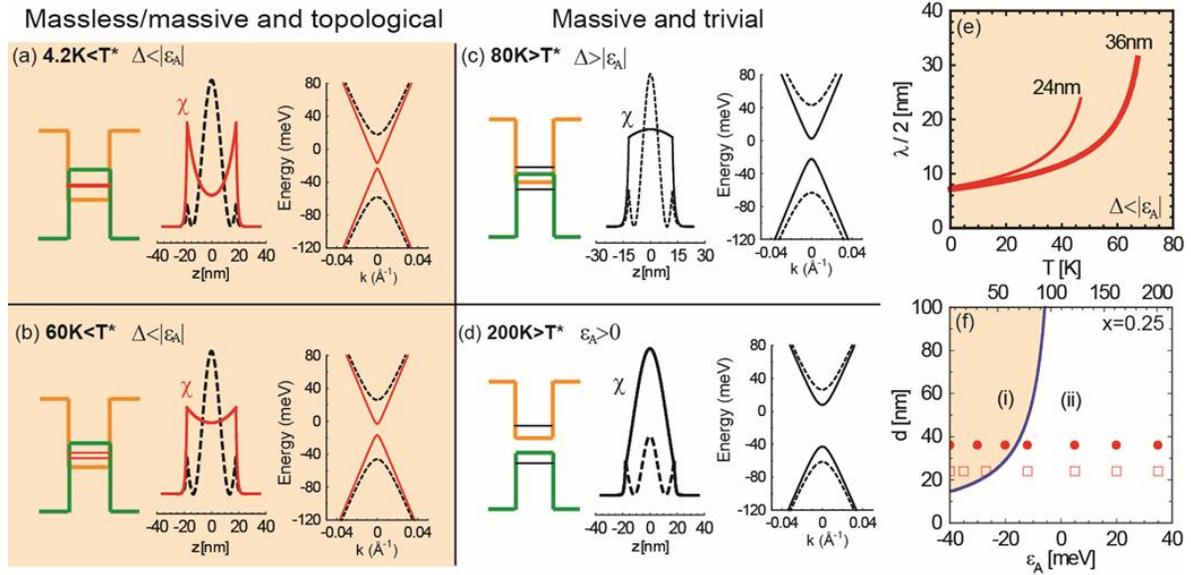

**FIG 4.** Results of k·p modeling of the data for TQW-36 at 4.2K (a), 60K (b), 80K (c) and 200K (d), showing for each temperature the quantum well band alignment with the position of the TQW ground states (red line), the probability density $\chi$ of the wave function of the TIS (E1) (red line) and the second QW subband E2 (dashed line), and their corresponding band dispersions (from left to right). (e) Penetration depth of the TIS versus temperature for the two investigated samples. (f) Phase diagram of the topological $Pb_{1-x}Sn_xSe$ quantum wells with x = 0.25 versus TQW thickness and temperature, respectively. Solid blue line represents the phase boundary between regime (i) and (ii). The experimental data points for the TQW-36 and TQW-24 are shown as full circles and empty squares respectively. The orange shading for (a,b,e and f) denotes the cases where $\Delta<|\varepsilon_A|$ (regime i) and the Dirac states are either massless

or massive but interfacial. In the other cases (c,d), $\Delta>|\varepsilon_A|$ and the massive Dirac states are no-longer pinned to the interface.

Lastly, we compute the continuous change of λ versus T, from Eq. (1) and the variation of $|\varepsilon_A|$ and Δ with temperature in regime (i). The results are displayed in Fig. 4(e), demonstrating the tunability of the phenomenon. Since the unique characteristics of the TIS such as their spin-momentum locking and scattering properties are linked to λ, Fig. 4(e) reveals how these can be continuously tuned in TCI TQWs. It is important to highlight here that the bulk Dirac cone of $Hg_{1-x}Cd_xTe$ has been also shown to be tunable with temperature. [49] However, in the case of HgTe, the 2D topological states and the heavy-hole band are known to hybridize. [50] [51] [52] Therefore, the main strength of $Pb_{1-x}Sn_xSe$, is that it hosts nearly-ideal Dirac cones with easily tunable characteristics.

### IV. Conclusion

*Overall, we have realized a TQW heterostructures based on heteroepitaxial $Pb_{1-x}Sn_xSe/Pb_{1-x}Eu_xSe$ superlattices and revealed the topological nature of the buried TIS using magnetooptical Landau level spectroscopy*. While previous works on topological thin films relied on surface sensitive probes that could not directly access the changing hybridization strength at subsurface topological interfaces, our optical technique is able to directly obtain this information via the changing energy gap of the interfacial Dirac states. It is thus important to recognize the importance of using such a technique to probe buried TIS in topological heterostructures. To this end, our results provide detailed phase diagrams (Figs. 3(f) and 4(f)) that demonstrate how the Dirac mass can be tuned by changing TQW thickness and more interestingly by varying temperature. *While this is a remarkable feature of the IV-VI TCI QW systems our conclusions are of general interest to other TI systems*. Keeping in mind the recently proposed analogy between the TQW and the 1D SSH model, [9] in that the hybridization of TIS in TQW can be related to hoping amplitude between lattice sites, the dynamics of the SSH model are temperature tunable in such a system.

Lastly, we highlight that our realization of high-quality superlattices and our successful observation of the TIS at TQW interfaces is a major and essential step towards engineering interfacial quantum phenomena predicted in TCIs. In fact, heterostructures of IV-VI materials with engineered lattice-mismatch, [53] [24] are predicted to host pseudo-Landau levels, helical flat-band states and an interfacial superconductor [25] [29] phase with an enhanced critical temperature. [28] Its possible coexistence with the TIS observed here in band gap engineered $Pb_{1-x}Sn_xSe$ or $Pb_{1-x}Sn_xTe$ TQW structures will provide a new platform to realize an intrinsic topological superconductor [54] for Majorana-based systems. We therefore motivate further work on strain and quantum engineering of TQW as a platform for emergent many-body topological states.

**Appendix A: Sample design and growth.**

Epitaxial growth of topological $Pb_{1-x}Sn_xSe/Pb_{1-y}Eu_ySe$ superlattices was performed using molecular beam epitaxy (MBE) in ultra-high vacuum of $5×10^{-10}$ mbar at a substrate temperature of 380°C on $BaF_2$ (111) using PbSe, SnSe, Eu, Te and $Bi_2Se_3$ effusion sources. The composition of the TQW $Pb_{1-x}Sn_xSe$ was set to x=25% to achieve band inversion, with a corresponding negative band gap $\varepsilon_A$ = -40 meV at 4.2K as checked on bulk-like reference layers. [33] $Pb_{1-y}Eu_ySe$ with x = 0.04 – 0.055 was chosen as barrier material. [36] Incorporation of Eu increases the band gap to a value of 270 – 314 meV (at 4.2K), as shown in the supplement. [33] This provides a stronger confinement of the TQW levels and pushes the absorption edge of the barriers well above the TQW LL transitions, while retaining a good lattice matching (0.6%). The thermal expansion coefficient of the well and the barrier are similar, [55] so the mismatch can be assumed constant with temperature.

To obtain a low carrier concentration of the TQWs, n-type doping using Bi was employed to compensate the native p-type character of $Pb_{1-x}Sn_xSe$ (see Ref. [32]) due to doubly charged cation vacancies. Superlattices were grown with the number of periods $N$ > 16 to obtain a sufficiently high absorption of the topological interface states in the infrared. Pseudomorphic 2D growth is achieved as checked by in-situ reflection high-energy electron diffraction. This yields a very high structural quality of the superlattices and interfaces as demonstrated by the x-ray diffraction spectra of Fig. 5(a,d), exhibiting a large number > 10 of sharp superlattice satellite peaks. The x-ray reciprocal space maps recorded around the symmetric (222) (Fig. 5(b,e)) and asymmetric (153) (Fig. 5(c,f)) Bragg reflection show that all satellite peaks are perfectly aligned along the [111] the growth direction, evidencing that the whole stack is fully pseudomorphic with coherent interfaces. Additionally, the observation of high order satellite peaks rules out any significant intermixing at the $Pb_{1-x}Sn_xSe$/ $Pb_{1-y}Eu_ySe$ interfaces. This is further confirmed by previous secondary ion mass spectroscopy measurements on similar superlattices shown in ref. [56]. More evidence of the excellent quality of the samples is provided in the supplementary material (S4) [33] where we estimate the sample mobility by analyzing the linewidth [57] of magnetooptical transitions. Mobilities of 15000 $cm^2$/Vs and 24000 $cm^2$/Vs are respectively found for TQW-24 and TQW-36.

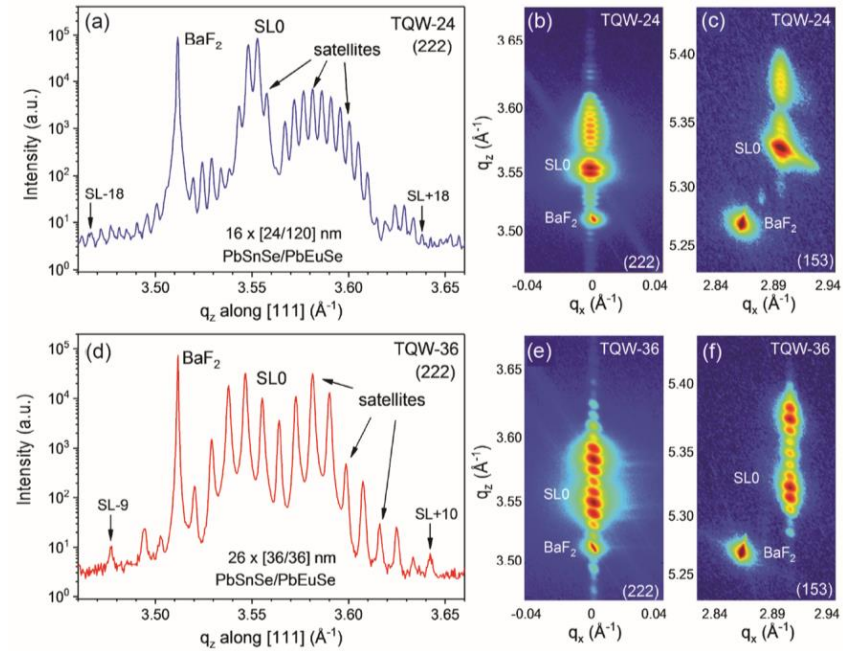

**FIG 5.** High Resolution x-ray diffraction data for the superlattice samples TQW24 (a-c) and TQW 36(d-f) taken using Cu-K$\alpha$1 radiation. (a,d) symmetric scans along the [111] growth direction. SL0 … SL±10/18 respectively denote superlattice satellite peaks stemming from the PbSnSe/PbEuSe stack, peak labeled BaF2 corresponds to the (222) Bragg reflection from the substrate. (b,e) reciprocal space maps depicting the scattered intensity distribution around the symmetric (222) reciprocal lattice point. (c,f) Same as (b,e) but around the asymmetric (153) reflection. Reciprocal lattice point $q_z$ denotes reciprocal space coordinate along [111], $q_x$ along an in-plane direction.

### Appendix B: k·p model to topological quantum wells

The topological quantum well system consisting of $Pb_{0.75}Sn_{0.25}Se$ band inverted quantum-wells and a non-inverted $Pb_{1-y}Eu_ySe$ barrier are modeled using the band profiles shown in Fig. 1. The $Pb_{1-y}Eu_ySe$ layer thickness is large enough to neglect coupling between the quantum wells across the barrier. For both materials, the band minima lie at the $L$-points of the Brillouin zone, the $L_6^-$ levels form the valence band and the $L_6^+$ level forms the conduction band, in $Pb_{0.75}Sn_{0.25}Se$ at 4.2K [46] [42] [58] while the opposite occurs in $Pb_{1-y}Eu_ySe$. For $Pb_{0.75}Sn_{0.25}Se$, this band-inversion is represented by a negative energy gap $\varepsilon_A < 0$ when the zero-energy reference is taken at the $L_6^+$ band edge in Fig. 1(c). In our model, we only take into account the interaction of the lowest conduction level and the highest valence level in the well material $Pb_{0.75}Sn_{0.25}Se$ (Fig. 1(c)). A 4-band $\boldsymbol{k.p}$ model that neglects the effect of far-bands is used to describe this system. In the basis $(|L_6^+, 1/2\rangle; |L_6^+, -1/2\rangle; |L_6^-, 1/2\rangle; |L_6^-, -1/2\rangle)$, for the topological case, the $\boldsymbol{k.p}$ Hamiltonian writes:

$$\begin{pmatrix} V_-(z) & 0 & \frac{P}{m_0}p_z & \frac{\hbar}{m_0}P(k_x - ik_y) \\ 0 & V_-(z) & \frac{\hbar}{m_0}P(k_x + ik_y) & -\frac{P}{m_0}p_z \\ \frac{P}{m_0}p_z^* & \frac{\hbar}{m_0}P(k_x - ik_y) & -|\varepsilon_A| + V_+(z) & 0 \\ \frac{\hbar}{m_0}P(k_x + ik_y) & -\frac{P}{m_0}p_z^* & 0 & -|\varepsilon_A| + V_+(z) \end{pmatrix} \quad (2)$$

where $m_0$ is the electron rest mass, $V_\pm(z)$ denotes the band offsets. It is 0 if $z < \left|\frac{d}{2}\right|$, and $V_\pm(z) = \pm V$ if not. $P$ is the Kane k·p matrix element. The band anisotropy is neglected as it is known to be very small (K≈1) in $Pb_{0.75}Sn_{0.25}Se$, and the four L-valleys are assumed equivalent. [32] In the Dirac formalism, we introduce $v_c = \frac{P}{m_0}$, the Dirac velocity.

V is assumed to be the same for the conduction and valence band, [36] yielding a symmetric confinement potential. This assumption holds for the low-energy subbands of interest to this work (H1, E1, H2, E2,...). Crystalline inversion symmetry is assumed to be preserved. In zone A of Fig. 1(c), the zero energy is taken at the $L_6^+$ band edge, so that the $L_6^-$ band edge is at $-|\varepsilon_A|$ at 4.2K. In zone B of Fig. 1(c), the $L_6^-$ band edge ends up being at $V - |\varepsilon_A|$ and the $L_6^+$ band edge at $-V$.

At $k_x = k_y = 0$, the problem is reduced to a two eigenvalues system:

$$\mathcal{H}\overrightarrow{\Psi_i} = E_i\overrightarrow{\Psi_i}$$

$$\mathcal{H}'\overrightarrow{\varphi_i} = E_i\overrightarrow{\varphi_i}$$

Where $\mathcal{H} = \begin{pmatrix} V_-(z) & -i\frac{\hbar}{m_0}P\frac{d}{dz} \\ -i\frac{\hbar}{m_0}P\frac{d}{dz} & -|\varepsilon_A| + V_+(z) \end{pmatrix}$, $\mathcal{H}' = \begin{pmatrix} V_-(z) & i\frac{\hbar}{m_0}P\frac{d}{dz} \\ i\frac{\hbar}{m_0}P\frac{d}{dz} & -|\varepsilon_A| + V_+(z) \end{pmatrix}$ and $\vec{\Psi_i} = \begin{pmatrix} F_1^{(i)} \\ F_2^{(i)} \end{pmatrix}$ and $\vec{\varphi_i} = \begin{pmatrix} F_1^{(i)} \\ -F_2^{(i)} \end{pmatrix}$.

$F_1^{(i)}$ and $F_2^{(i)}$ are respectively the $L_6^+$ and the $L_6^-$ component of the envelope wavefunction of the $i$-th bound states.

We use the classical probability current continuity conditions for $F_1^{(i)}$ to solve the problem [1]:

$$F_1^{(i)} \text{ continuous at } |z| = \frac{d}{2}$$

$$\frac{1}{-|\varepsilon_A| + V_+(z) - E_i} \frac{dF_1^{(i)}}{dz} \text{ continuous at } |z| = \frac{d}{2}$$

The term $\frac{1}{-|\varepsilon_A|+V_+(z)-E_i}$ is equivalent to an inverse of an effective mass in a non-parabolic (Dirac) system. It changes sign at the interfaces. A wavefunction of the form $F_1^{(i)} = A_i \cos(kz)$ or $F_1^{(i)} = B_i \sin(kz)$, is used and is referred to as the even and odd case, respectively.

The continuity conditions applied to $F_1^{(i)}$ yields a system of two transcendental equations that can be solved to find the energy eigenvalue and the eigenfunction:

For E>0 and E<-|ε_A|

$$\tan\left(\frac{kd}{2}\right) = \frac{\rho}{k}\frac{E+|\varepsilon_A|}{E+|\varepsilon_A|-V} \quad \text{even case}$$

$$\cotan\left(\frac{kd}{2}\right) = -\frac{\rho}{k}\frac{E+|\varepsilon_A|}{E+|\varepsilon_A|-V} \quad \text{odd case}$$

Here: $k = \sqrt{\frac{2m_A E}{\hbar^2}\left(1+\frac{E}{|\varepsilon_A|}\right)}$, $\rho = \sqrt{\frac{2m_A}{\varepsilon_A \hbar^2}(E+V)(-E-|\varepsilon_A|+V)}$ and $m_A = \frac{|\varepsilon_A|}{2v_c^2} > 0$

For -|ε_A|<E<0

$$\tanh\left(\frac{\kappa d}{2}\right) = -\frac{\rho}{\kappa}\frac{E+|\varepsilon_A|}{E+|\varepsilon_A|-V} \quad \text{even case}$$

$$\cotanh\left(\frac{\kappa d}{2}\right) = -\frac{\rho}{\kappa}\frac{E+|\varepsilon_A|}{E+|\varepsilon_A|-V} \quad \text{odd case}$$

Here: $\kappa = \sqrt{-\frac{2m_A E}{\hbar^2}\left(1+\frac{E}{|\varepsilon_A|}\right)} = \sqrt{-\frac{|\varepsilon_A|E}{\hbar^2 v_c^2}\left(1+\frac{E}{|\varepsilon_A|}\right)} = \frac{1}{\hbar v_c}\sqrt{-E(|\varepsilon_A|+E)}$

In this case, the change of topology at the interfaces causes the wave vector $k$ to be imaginary $k = i\kappa$, therefore yielding $F_1^{(1)} = A\cosh(\kappa z)$ for -|ε_A|<E<0. The probability density of the first subband $\chi(z) = |F_1^{(1)}|^2$ is therefore

peaked at the interface. It is the topological interface state (TIS) of the QW. Its penetration depth into the well is therefore given by:

$$\lambda = \frac{1}{\kappa}$$

The TIS energy eigenvalues can be expressed as

$$E = \frac{-|\varepsilon_A|}{2} \pm \frac{\Delta}{2}$$

The expression for the penetration depth is therefore reduced to:

$$\lambda = \frac{2\hbar v_c}{\sqrt{\varepsilon_A^2 - \Delta^2}}$$

$\chi(z)$ is shown in the inset of Fig. 2(c) for TQW-36. In general, the method used here is equivalent to a Ben-Daniel Duke model where the non-parabolicity ('Dirac-ness') of the energy bands is accounted for. This method is discussed thoroughly in Ref. [1], [34] and [35]. It is similar to the methods used in previous works to compute the effect of confinement of the topological surface states in TIs and TCIs. [27], [47], [48]

### Appendix C: Derivation of the k-dispersion and Landau levels.

We derive an effective Hamiltonian using the $k_{x,y}=0$ solutions found by solving the transcendental equations. The effective Hamiltonian is expressed in the orthonormal $(\overrightarrow{\Psi_n}, \overrightarrow{\varphi_n})_{0<n<i}$ basis. The $k_x$, $k_y$-dependent part of the 4x4 $\mathbf{k}.\mathbf{p}$ Hamiltonian (Eq. 2) is regarded as a perturbation. We can write it as:

$$H_{mn}^{eff}(k_x, k_y) = \int_{-\infty}^{+\infty} dz \langle \overrightarrow{\Psi_m}, \overrightarrow{\varphi_m}|\Delta H|\overrightarrow{\Psi_n}, \overrightarrow{\varphi_n}\rangle$$

With,

$$\Delta H = \begin{pmatrix} 0 & 0 & 0 & \frac{\hbar}{m_0}P(k_x - ik_y) \\ 0 & 0 & \frac{\hbar}{m_0}P(k_x + ik_y) & 0 \\ 0 & \frac{\hbar}{m_0}P(k_x - ik_y) & 0 & 0 \\ \frac{\hbar}{m_0}P(k_x + ik_y) & 0 & 0 & 0 \end{pmatrix}$$

The eigenvalue problem corresponding to the effective Hamiltonian is then numerically solved for each $k_{x,y}$ in the range $0 < k_{x,y} < 0.05$ Å$^{-1}$.

The Landau levels calculation follows exactly the same method. The perturbation is expressed in terms of the magnetic field. With $N$ denoting the Landau index, it is written as:

$$\Delta H = \begin{pmatrix} 0 & 0 & 0 & \sqrt{2e\hbar v_c^2 B(N+1)} \\ 0 & 0 & \sqrt{2e\hbar v_c^2 B(N+1)} & 0 \\ 0 & \sqrt{2e\hbar v_c^2 B(N+1)} & 0 & 0 \\ \sqrt{2e\hbar v_c^2 B(N+1)} & 0 & 0 & 0 \end{pmatrix}$$

The eigenvalue problem corresponding to the effective B-dependent Hamiltonian is numerically solved for 0≤B≤15T, and -1≤$N$≤10.

**Appendix D: Anti-crossing of Landau levels.**

An anti-crossing between $E_{i,N}$ and $E_{i\pm1,N\pm1}$ Landau levels (i: band index, $N$: LL index) is computed using the conventional approach of solving the eigenvalue problem corresponding to the interaction of the two levels according to:

$$\begin{pmatrix} E_{i,N} & W \\ W & E_{i\pm1,N\pm1} \end{pmatrix}$$

Here $W$ is magnitude of the avoided crossing. It is taken to be equal to 5meV for the -1$^{E1}$ 0$^{E2}$ anticrossing shown in Fig. 2(b,d). $W$ decreases significantly with increasing $N$. This type of anticrossing is characteristic of the Landau levels of non-parabolic semiconductors with band minima at the L-points. [59]

**Acknowledgments.** We acknowledge fruitful discussions with D. Heiman. This work is supported by Agence Nationale de la Recherche grant ENS-ICFP Grant No. ANR-10-LABX-0010/ANR-10-IDEX-0001-02 PSL and by the Austrian Science Fund, Projects P28185-N27 and P29630-N27.. G.K. is partly supported by a PSL PhD thesis scholarship.

[1]   G. Bastard, *Wave Mechanics Applied to Semiconductor Heterostructures* (Les éditions de physique, Les Ulis, France, 1996).

[2]   M. Konig, S. Wiedmann, C. Brune, A. Roth, H. Buhmann, L. W. Molenkamp, X.-L. Qi, and S.-C. Zhang, Science (80-. ). **318**, 766 (2007).

[3]   B. Hunt, J. D. Sanchez-Yamagishi, A. F. Young, M. Yankowitz, B. J. LeRoy, K. Watanabe, T. Taniguchi, P. Moon, M. Koshino, P. Jarillo-Herrero, and R. C. Ashoori, Science (80-. ). **340**, 1427 (2013).

[4]   J. Faist, F. Capasso, D. L. Sivco, C. Sirtori, A. L. Hutchinson, and A. Y. Cho, Science (80-. ). **264**, 553 (1994).

[5]   R. M. Stevenson, R. J. Young, P. Atkinson, K. Cooper, D. A. Ritchie, and A. J. Shields, Nature **439**, 178 (2006).

[6]   M. Z. Hasan and C. L. Kane, Rev. Mod. Phys. **82**, 3045 (2010).

[7]   X.-L. Qi and S.-C. Zhang, Rev. Mod. Phys. **83**, 1057 (2011).

[8]   a. a. Burkov and L. Balents, Phys. Rev. Lett. **107**, 127205 (2011).

[9]   I. Belopolski, S. Y. Xu, N. Koirala, C. Liu, G. Bian, V. N. Strocov, G. Chang, M. Neupane, N. Alidoust, D. Sanchez, H. Zheng, M. Brahlek, V. Rogalev, T. Kim, N. C. Plumb, C. Chen, F. Bertran, P. Le Fèvre, A. Taleb-Ibrahimi, M. C. Asensio, M. Shi, H. Lin, M. Hoesch, S. Oh, and M. Z. Hasan, Sci. Adv. **3**, 12 (2017).

[10]  L. Fu, Phys. Rev. Lett. **106**, 106802 (2011).

[11]  T. H. Hsieh, H. Lin, J. Liu, W. Duan, A. Bansil, and L. Fu, Nat. Commun. **3**, 982 (2012).

[12]  Y. Ando and L. Fu, Annu. Rev. Condens. Matter Phys. **6**, 361 (2015).

[13]  P. Dziawa, B. J. Kowalski, K. Dybko, R. Buczko, A. Szczerbakow, M. Szot, E. Łusakowska, T. Balasubramanian, B. M. Wojek, M. H. Berntsen, O. Tjernberg, and T. Story, Nat. Mater. **11**, 1023 (2012).

[14]  P. S. Mandal, G. Springholz, V. V. Volobuev, O. Caha, A. Varykhalov, E. Golias, G. Bauer, O. Rader, and J. Sánchez-Barriga, Nat. Commun. **8**, 968 (2017).

[15]  S.-Y. Xu, C. Liu, N. Alidoust, M. Neupane, D. Qian, I. Belopolski, J. D. D. Denlinger, Y. J. J. Wang, H. Lin, L. A. a. Wray, G. Landolt, B. Slomski, J. H. H. Dil, A. Marcinkova, E. Morosan, Q. Gibson, R. Sankar, F. C. C. Chou, R. J. J. Cava, a. Bansil, and M. Z. Z. Hasan, Nat. Commun. **3**, 1192 (2012).

[16]  C. Yan, J. Liu, Y. Zang, J. Wang, Z. Wang, P. Wang, Z.-D. Zhang, L. Wang, X. Ma, S. Ji, K. He, L. Fu, W. Duan, Q.-K. Xue, and X. Chen, Phys. Rev. Lett. **112**, 186801 (2014).

[17]  C. Fang, M. J. Gilbert, and B. A. Bernevig, Phys. Rev. Lett. **112**, 46801 (2014).

[18]  J. Liu, T. H. Hsieh, P. Wei, W. Duan, J. Moodera, and L. Fu, Nat. Mater. **13**, 178 (2014).

[19]  V. V Volobuev, P. S. Mandal, M. Galicka, O. Caha, J. Sánchez-Barriga, D. Di Sante, A. Varykhalov, A. Khiar, S. Picozzi, G. Bauer, P. Kacman, R. Buczko, O. Rader, and G. Springholz, Adv. Mater. **29**, 1604185 (2017).

[20]  I. Sodemann, Z. Zhu, and L. Fu, Phys. Rev. X **7**, 1 (2017).

[21]  Y. Okada, M. Serbyn, H. Lin, D. Walkup, W. Zhou, C. Dhital, M. Neupane, S. Xu, Y. J. Wang, R. Sankar, F. Chou, A. Bansil, M. Z. Hasan, S. D. Wilson, L. Fu, and V. Madhavan, Science **341**, 1496 (2013).

[22]  I. Pletikosić, G. D. Gu, and T. Valla, Phys. Rev. Lett. **112**, 146403 (2014).

[23]  C. M. Polley, P. Dziawa, A. Reszka, A. Szczerbakow, R. Minikayev, J. Z. Domagala, S. Safaei, P. Kacman, R. Buczko, J. Adell, M. H. Berntsen, B. M. Wojek, O. Tjernberg, B. J. Kowalski, T. Story, and T. Balasubramanian, Phys. Rev. B - Condens. Matter Mater. Phys. **89**, 75317 (2014).

[24]  I. Zeljkovic, D. Walkup, B. A. Assaf, K. L. Scipioni, R. Sankar, F. Chou, and V. Madhavan, Nat. Nanotechnol. **10**, 849 (2015).

[25]  E. Tang and L. Fu, Nat Phys **10**, 964 (2014).

[26]  B. A. Assaf, T. Phuphachong, E. Kampert, V. V. Volobuev, P. S. Mandal, J. Sánchez-Barriga, O. Rader, G. Bauer, G. Springholz, L. A. De Vaulchier, and Y. Guldner, Phys. Rev. Lett. **119**, 106602 (2017).

[27]  J. Liu and L. Fu, Phys. Rev. B **91**, 81407 (2015).

[28]  G. E. Volovik, Phys. Scr. **T164**, 14014 (2015).

[29]  V. J. Kauppila, F. Aikebaier, and T. T. Heikkilä, Phys. Rev. B **93**, 214505 (2016).

[30] Y. Wang, G. Luo, J. Liu, R. Sankar, N.-L. Wang, F. Chou, L. Fu, and Z. Li, Nat. Commun. **8**, 366 (2017).

[31] B. A. Assaf, T. Phuphachong, V. V. Volobuev, A. Inhofer, G. Bauer, G. Springholz, L. A. de Vaulchier, and Y. Guldner, Sci. Rep. **6**, 20323 (2016).

[32] B. A. Assaf, T. Phuphachong, V. V Volobuev, G. Bauer, G. Springholz, L.-A. De Vaulchier, and Y. Guldner, NPJ Quantum Mater. **2**, 26 (2017).

[33] S. supplementary material for detail on growth; T. studies; H. effect measurements; a discussion of ad hoc explanations to our results; measurements of the resistance as function H. and details on the model used to simulate the magnetic domain Dynamics., (n.d.).

[34] G. Bastard, Phys. Rev. B **24**, 5693 (1981).

[35] Y.-C. Chang, J. N. Schulman, G. Bastard, Y. Guldner, and M. Voos, Phys. Rev. B **31**, 2557 (1985).

[36] M. Simma, G. Bauer, and G. Springholz, Appl. Phys. Lett. **101**, 172106 (2012).

[37] M. Simma, T. Fromherz, G. Bauer, and G. Springholz, Appl. Phys. Lett. **95**, 212103 (2009).

[38] G. Bastard, C. Rigaux, Y. Guldner, J. Mycielski, and A. Mycielski, J. Phys. Fr. **39**, 87 (1978).

[39] G. Bauer, H. Pascher, and W. Zawadzki, Semicond. Sci. Technol. **7**, 703 (1992).

[40] M. Neupane, A. Richardella, J. Sánchez-Barriga, S. Xu, N. Alidoust, I. Belopolski, C. Liu, G. Bian, D. Zhang, D. Marchenko, A. Varykhalov, O. Rader, M. Leandersson, T. Balasubramanian, T.-R. Chang, H.-T. Jeng, S. Basak, H. Lin, A. Bansil, N. Samarth, and M. Z. Hasan, Nat. Commun. **5**, 3841 (2014).

[41] K. He, Y. Zhang, K. He, C. Chang, C. Song, L. Wang, X. Chen, J. Jia, Z. Fang, X. Dai, W. Shan, S. Shen, Q. Niu, X. Qi, S. Zhang, X.-C. Ma, and Q.-K. Xue, Nat. Phys. **6**, 584 (2010).

[42] G. Bauer, in *Narrow Gap Semicond. Phys. Appl. Proceeding Int. Summer Sch.*, edited by W. Zawadzki (Springer Berlin Heidelberg, Berlin, Heidelberg, 1980), pp. 427–446.

[43] Y. Tanaka, T. Sato, K. Nakayama, S. Souma, T. Takahashi, Z. Ren, M. Novak, K. Segawa, and Y. Ando, Phys. Rev. B - Condens. Matter Mater. Phys. **87**, 1 (2013).

[44] T. Liang, Q. Gibson, J. Xiong, M. Hirschberger, S. P. Koduvayur, R. J. Cava, and N. P. Ong, Nat. Commun. **4**, 3696 (2013).

[45] B. M. Wojek, P. Dziawa, B. J. Kowalski, A. Szczerbakow, a. M. Black-Schaffer, M. H. Berntsen, T. Balasubramanian, T. Story, and O. Tjernberg, Phys. Rev. B **90**, 161202 (2014).

[46] A. J. Strauss, Phys. Rev. **157**, 608 (1967).

[47] R. Buczko and Ł. Cywiński, **85**, 205319.

[48] H.-Z. Lu, W.-Y. Shan, W. Yao, Q. Niu, and S.-Q. Shen, Phys. Rev. B **81**, 115407 (2010).

[49]   F. Teppe, M. Marcinkiewicz, S. S. Krishtopenko, S. Ruffenach, C. Consejo,  a M. Kadykov, W. Desrat, D. But, W. Knap, J. Ludwig, S. Moon, D. Smirnov, M. Orlita, Z. Jiang, S. V. Morozov, V. I. Gavrilenko, N. N. Mikhailov, and S. A. Dvoretskii, Nat. Commun. **7**, 12576 (2016).

[50]   V. Dziom, A. Shuvaev, A. Pimenov, G. V. Astakhov, C. Ames, K. Bendias, J. Böttcher, G. Tkachov, E. M. Hankiewicz, C. Brüne, H. Buhmann, and L. W. Molenkamp, Nat. Commun. **8**, 15197 (2017).

[51]   A. Inhofer, S. Tchoumakov, B. A. Assaf, G. Fève, J. M. Berroir, V. Jouffrey, D. Carpentier, M. O. Goerbig, B. Plaçais, K. Bendias, D. M. Mahler, E. Bocquillon, R. Schlereth, C. Brüne, H. Buhmann, and L. W. Molenkamp, Phys. Rev. B **96**, (2017).

[52]   C. Brüne, C. Thienel, M. Stuiber, J. Böttcher, H. Buhmann, E. G. Novik, C.-X. Liu, E. M. Hankiewicz, and L. W. Molenkamp, Phys. Rev. X **4**, 41045 (2014).

[53]   G. Springholz and K. Wiesauer, 1 (2002).

[54]   F. Zhang, C. L. Kane, and E. J. Mele, Phys. Rev. Lett. **111**, 56403 (2013).

[55]   A. Katzir, R. Rosman, Y. Shani, K. H. Bachem, H. Böttner, and H. Preier, in *Handb. Solid-State Lasers*, edited by P. Cheo (1989).

[56]   G. Springholz, A. Holzinger, H. Krenn, H. Clemens, G. Bauer, H. Böttner, P. Norton, and M. Maier, J. Cryst. Growth **113**, 593 (1991).

[57]   T. Ando and Y. Uemura, J. Phys. Soc. Japan **36**, 959 (1974).

[58]   D. L. Mitchell and R. F. Wallis, Phys. Rev. **151**, 581 (1966).

[59]   D. C. Tsui, G. Kaminsky, and P. H. Schmidt, Phys. Rev. B **9**, 3524 (1974).